\documentclass[a4paper, 11pt]{article}
\usepackage{comment} 
\usepackage[nointegrals]{wasysym}
\usepackage{fullpage} 
\usepackage{graphicx}
\usepackage{cite}
\usepackage{bm}
\usepackage{float}
\usepackage{amsmath}
\usepackage[utf8]{inputenc}
\usepackage{amssymb}
\usepackage{mathrsfs}
\usepackage{wrapfig}
\usepackage{enumitem}
\usepackage[english]{babel}
\usepackage{comment}
\usepackage{bbold}  
\usepackage{multicol}
\usepackage[usenames,dvipsnames]{xcolor} 
\usepackage{xcolor,hyperref}
\usepackage{setspace}
\usepackage{abstract}
\usepackage{sectsty}
\sectionfont{\large}
\usepackage{tablefootnote}

\usepackage[T1]{fontenc}
\usepackage[margin=0.94in]{geometry}
\begin{document}
\vspace*{\fill}
\begin{center}
    \Large\textbf{\textsc{ \textcolor{Black}{Unveiling the inconsistency of the Proca theory\\ with non-minimal coupling to gravity} }}
    
    \normalsize\textsc{Anamaria Hell$^{1}$}
\end{center}

\begin{center}
    $^{1}$ \textit{Kavli IPMU (WPI), UTIAS,\\ The University of Tokyo,\\ Kashiwa, Chiba 277-8583, Japan}
    \\ \href{mailto:anamaria.hell@ipmu.jp}{anamaria.hell@ipmu.jp} 
\end{center}
\thispagestyle{empty} 

\renewcommand{\abstractname}{\textsc{\textcolor{Black}{Abstract}}}

\begin{abstract}
 We study the degrees of freedom of the Proca theory, non-minimally coupled to gravity. In the Minkowski background, this theory propagates five degrees of freedom -- a massive longitudinal mode, two massive vector ones, and two massless tensor modes. At first sight, the non-linear coupling between the metric perturbations and the vector field indicates that both longitudinal and tensor modes become strongly coupled, at the same scale. This would imply that no matter how small the photon mass is if non-minimal coupling is taken into account, gravitational waves would necessarily be strongly coupled. We show that the way out of this inconsistency is through the introduction of the disformal coupling to the metric perturbations that resemble the vector-type disformal transformations. This way, the unphysical coupling between the two types of modes can be avoided, rendering the model consistent. As a result, we show that only the longitudinal modes enter a strong coupling regime, while both tensor and transverse modes remain weakly coupled at all scales up to the Planck length. Finally, using the same form of the disformal transformation, we introduce a disformal frame in which the recently reported runaway modes are absent.
  \end{abstract}
 
\vfill
\clearpage
\pagenumbering{arabic} 
\newpage

\section{  {\textcolor{Black}{\Large \textbf{\textsc{Introduction}}}}}

One of the most important contemporary puzzles underlying the structure of our Universe is -- \textit{What is dark matter?} The existence of this component, which should account for the \textit{missing mass} of the Universe is strongly supported by the observational data. Yet, the theory that explains its origin remains unknown. 

Among numerous proposals for models of dark matter, the theory of the dark photon is particularly interesting. This theory assumes that the photon is described by a massive and stable vector field, and can be detectable due to the kinetic mixing with the ordinary photon \cite{Fabbrichesi:2020wbt, Caputo:2021eaa}. Because of this mixing, these dark matter particles can be produced in the early Universe, for a sufficiently large mass. \textit{Gravitational particle production} -- the mechanism in which the particles are produced due to the curvature of the space-time -- can account for their production in the early Universe if, in contrast, the photon is ultra-light \cite{ Kolb:2023ydq}. To date, this mechanism has brought to us the understanding of the origin of galaxies and the large-scale structure, tracing its seeds back to the quantum fluctuations during an inflationary era \cite{Mukhanov:1981xt, Chibisov:1982nx, Mukhanov:1990me, Mukhanov:1988jd, Kodama:1984ziu, Sasaki:1986hm, Sasaki:1983kd}, and is even responsible for the Hawking radiation of black holes \cite{Hawking:1974rv}. 
Notably, with it, the dark photons can be produced independently of any coupling to the standard-model particles, making the particles \textit{completely dark} \cite{Kolb:2020fwh}.  

The theory underlying the physics of these dark matter particles is known as the Proca theory \cite{Proca:1936fbw}. When considered in curved space-time, one way to study it is by assuming the minimal coupling between the vector field and the curvature, through a simple generalization of the volume element in the action and promotion of the partial derivatives to the covariant ones. The cosmological implications of this theory were widely studied in \cite{Graham:2015rva, Ahmed:2020fhc, Demozzi:2009fu}.  
Recently, the possibility of having non-minimal couplings between a vector field and the curvature tensor has become very attractive as well.  This natural extension of the model, whose quantization was discussed in \cite{Buchbinder:2017zaa, Toms:2015fja}, was particularly intriguing due to its possibility to realize inflation due to a vector field and generate the primordial magnetic fields \cite{Golovnev:2008cf, Himmetoglu:2008zp, Turner:1987bw, Dimopoulos:2008rf, Dimopoulos:2011ws}. Moreover, it can influence the gravitational particle production of the dark photon \cite{Cembranos:2023qph, Kolb:2020fwh, Ozsoy:2023gnl, Capanelli:2024pzd}. In this work, we will also consider Proca theory, non-minimally coupled to gravity, studying it from a complementary perspective that arises from the recent developments in massive gauge theories. 

A natural question that arises when studying massive gauge theories is -- \textit{How do they behave when the mass is set to zero?} According to the principle of continuity, in such a limit, all observables should match those of the corresponding massless theory \cite{BassSCh1995}.   Interestingly, however, the first insights into such question posed non-trivial challenges, originally noticed in massive Yang-Mills theory (mYM) with mass added \textit{by hand}, and massive linearized gravity \cite{Fierz:1939ix, Glashow:1961tr}. Studying the two with conventional methods could lead to the conclusion that they might not be realized in nature. In the case of mYM theory, when calculating the corrections to the propagator of the non-Abelian vector field it was shown that the perturbative series are singular in mass \cite{vanDam:1970vg, Reiff:1969pq, Slavnov:1972qb, Veltman:1970dwc}. For the mGR case, the situation was even more drastic. It yielded predictions that were different from those of Einstein's relativity and remained such in the massless limit. 

As pointed out in \cite{vanDam:1970vg, Zakharov:1970cc}, the occurrence of this perturbative discontinuity in mass was due to the appearance of the extra \textit{degrees of freedom (dof)}, the longitudinal modes, that failed to decouple from the remaining dof in the massless limit. However, in \cite{Vainshtein:1972sx}, it was noticed in the context of massive gravity that once the non-linear terms become of the same order as the linear ones, the perturbation theory for the longitudinal mode breaks down. Subsequently, on smaller distances than this Vainshtein radius, the longitudinal modes are strongly coupled, and one obtains predictions of linearized gravity up to small corrections \cite{Deffayet:2001uk, Gruzinov:2001hp}. The possibility of the smooth massless limit for the mYM was conjectured in \cite{Vainshtein:1971ip} as well, and later shown in \cite{Hell:2021oea}.  \textit{The Vainsthein mechanism} -- mechanism which is responsible for this behavior of the modes absent in massless theories was also generalized to other field theories \cite{ Dvali:2006su, Dvali:2007kt, Chamseddine:2010ub, Alberte:2010it, Mukohyama:2010xz, deRham:2010ik, deRham:2010kj, Chamseddine:2012gh, Tasinato:2014eka, Chamseddine:2018gqh, Hell:2021wzm, Hell:2023mph, Heisenberg:2014rta, Heisenberg:2020xak,  BeltranJimenez:2016rff}.

Similarly to the mGR and mYM theories, Proca theory with cubic and quartic self-interactions obeys the Vainshtein mechanism as well - the longitudinal mode that is absent in the massless theory indicates a discontinuity in the perturbative series, becomes strongly coupled once this series breaks down and decouples from the remaining dof beyond such scales, which in turn remain weakly coupled \cite{Dvali:2007kt, Hell:2021wzm}. 
In this paper, we will also explore the massless limit of the Proca theory, now focusing on its non-minimal coupling to gravity, through the Ricci scalar and the Ricci tensor. 
In particular, motivated by the previous studies on Proca theory, and massive gauge theories in general, we will aim to answer -- \textit{Does Proca theory with non-minimal coupling to gravity have strong coupling in flat spacetime, and if yes, how strongly coupled the theory is? }

Based on the results of massive gauge theories, there are three possible outcomes: First, the theory could be already well-behaved in the massless limit, despite the presence of the longitudinal modes. 
The theory might likewise contain perturbative series that are singular in mass, which would consequently imply strong coupling. As the longitudinal modes are absent in the massless theory, it would be natural to expect that they become strongly coupled, with the rest of the modes remaining in the weakly coupled regime. This has been shown in the presence of cubic or quartic self-interactions. Another possibility is that the remaining modes enter the strong coupling regime as well. 
 This outcome, in which the transverse modes, or tensor modes -- modes of the metric perturbations -- become strongly coupled, however,  would be unphysical.  It would indicate that in the presence of non-minimal coupling, one could either not be able to recover electromagnetism or discover gravitational waves. 

By studying the Proca theory, non-minimally coupled to gravity in a flat background, and taking into account the metric perturbations around it, we will show that, surprisingly, the last case takes place. Similarly to mYM theory, both types of coupling will indicate a perturbative series that is singular in mass. At the same time, following the close analogy to the Vainshtein mechanism, due to the non-linear terms, the perturbation theory will break down at the corresponding scales. As we will see, while the Ricci scalar coupling introduces a natural strong coupling that involves only the longitudinal modes, the coupling of the Proca field to the Ricci tensor introduces a strong coupling in both longitudinal and tensor modes, thus yielding in a physically inconsistent result.

Moreover, we will show that the way to resolve this ambiguity lies in the introduction of the disformal coupling to the metric perturbations simultaneously. Such disformal transformations, which modify the original metric by adding a bilinear term in the external fields or their field strength have gained a lot of attention in cosmology, and construction of new theories of gravity \cite{Bekenstein:1992pj, Gumrukcuoglu:2019ebp, Deffayet:2013tca, Kimura:2016rzw, Domenech:2018vqj, DeFelice:2019hxb, Jirousek:2022rym, Takahashi:2021ttd, Deruelle:2014zza, Domenech:2023ryc, Alinea:2022ygr, Naruko:2019jzj, Heisenberg:2016eld, Papadopoulos:2017xxx, Bettoni:2013diz, Zumalacarregui:2013pma, Domenech:2015hka, Domenech:2015tca, Crisostomi:2016czh, BenAchour:2016cay}. As we will see, in flat spacetime, the background value for the Ricci tensor is vanishing, so the form of the disformal coupling necessary to remove the tensor mode ambiguity will be possible to deduct by studying the form of the coupling already on the level of action. In this paper, we will also infer and confirm its form by closely following the analogy with the mYM theory.

Recently, in \cite{Capanelli:2024pzd} it was reported that the Proca theory with non-minimal coupling to gravity suffers from another problem in the curved background -- the production of runaway modes which have a negative frequency. In this work, we will show that the origin of this instability lies due to the coupling of the vector field with the Ricci tensor. While generically the instability might persist, we will show that in the disformal frame, defined by the same form of the disformal transformation as in the flat background, and in which the matter is minimally coupled to gravity, the runaway modes are absent.

The paper is organized in the following way: In section 2, we will review the free Proca theory and linearized gravity in a flat space-time background. In section 3, we will study the massless limit of the Proca theory with non-minimal couplings to gravity. By treating the Ricci scalar and Ricci tensor couplings separately, we will show the appearance of unphysical behavior of the tensor and longitudinal modes in the case of the latter interaction. Following the close analogy with the mYM theory, we will introduce the resolution to this puzzle in section 4.  This will be followed by the resolution of the runaway in the curved space-time in section 5. Finally, we will conclude with a discussion of the results and prospects of the theory.

\section{  {\textcolor{Black}{\Large \textbf{\textsc{The basics of linearized gravity and Proca theory}}}}}
The Proca theory, the theory of a massive vector field, which is non-minimally coupled to gravity, is described by the following action\footnote{We use the metric signature (-,+,+,+).}: 
\begin{equation}\label{PdeBroglieAction}
    S=S_{EH}+S_{P}+S_{nmin},
\end{equation}
where
\begin{equation*}
    \begin{split}
        &S_{EH}=\frac{M_{pl}^2}{2}\int d^4x\sqrt{-g}R\\
        &S_{P}=\int d^4x \sqrt{-g}\left(-\frac{1}{4}g^{\mu\alpha}g^{\nu\beta}F_{\mu\nu}F_{\alpha\beta}-\frac{m^2}{2}g^{\mu\nu}A_{\mu}A_{\nu}\right)\\
        &S_{nmin}=-\frac{1}{2}\int d^4x \sqrt{-g}\left(\alpha  Rg^{\mu\nu}A_{\mu}A_{\nu}+\beta R^{\mu\nu}A_{\mu}A_{\nu}\right).
    \end{split}
\end{equation*}

The goal of this section is to analyze the degrees of freedom \textit{dof} of this theory, starting with the flat background. For this, let us perturb the metric in the following way:  
\begin{equation}
    g_{\mu\nu}=\eta_{\mu\nu}+h_{\mu\nu},
\end{equation}
where $\eta_{\mu\nu}$ is the Minkowski metric. Then,  we find that the Lagrangian density corresponding to (\ref{PdeBroglieAction}) at linearized level is given by: 
\begin{equation}
    \mathcal{L}=\mathcal{L}_{EH}+\mathcal{L}_{P},
\end{equation}
where 
\begin{equation}\label{linearized}
\begin{split}
    \mathcal{L}_{EH}=&\frac{M_{pl}^2}{4}\left(-\partial_{\alpha}h^{\mu\nu}\partial^{\alpha}h_{\mu\nu}+2\partial_{\mu}h^{\mu\nu}\partial_{\alpha}h_{\nu}^{\alpha}-2\partial_{\nu}h\partial_{\mu}h^{\mu\nu}+\partial_{\mu}h\partial^{\mu}h\right),\\
    \mathcal{L}_{P}=&-\frac{1}{4}F_{\mu\nu}F^{\mu\nu}-\frac{m^2}{2}A_{\mu}A^{\mu}. 
    \end{split}
\end{equation}
Here, the indices are raised and lowered with the Minkowski metric. 

Let us now first study the degrees of freedom \textit{(dof)} in gravity. For this, we will decompose the metric perturbations according to the spatial rotations: 
\begin{equation}\label{decomposition}
    \begin{split}
        &h_{00}=2\phi\\
        &h_{0i}=B_{,i}+S_i,\qquad\qquad S_{i,i}=0\\
        &h_{ij}=2\psi\delta_{ij}+2E_{,ij}+F_{i,j}+F_{j,i}+h_{ij}^{T},\qquad\qquad F_{i,i}=0,\quad h_{ij,i}^{T}=0,\quad h_{ii}^{T}=0,
    \end{split}
\end{equation}
where $,i=\frac{\partial}{\partial x^i}$. 
Then, the gravitational contribution of (\ref{linearized}) becomes: 
\begin{equation}
    \begin{split}
\mathcal{L}_{EH}=&M_{pl}^2\left(2\phi\Delta\psi-3\dot{\psi}\dot{\psi}-\psi\Delta\psi-2\psi\Delta\dot{\sigma}\right)-\frac{M_{pl}^2}{4}V_i\Delta V_i-\frac{M_{pl}^2}{8}\partial_{\mu}h_{ij}^T\partial^{\mu}h_{ij}^T,
     \end{split}
\end{equation}
where 
\begin{equation}
    \sigma=B-\dot E,\qquad \text{and}\qquad V_i=S_i-\dot F_i.
\end{equation}
In the above relations, the dot denotes a derivative with respect to the time. We can notice that $\phi$ and $S_i$ are not propagating. By varying the action with respect to them, we find the following equations respectively: 
\begin{equation}
     \Delta\psi=0\qquad\text{and}\qquad \Delta V_i=0,
\end{equation}
whose solutions are given by: 
\begin{equation}
    \psi=0\qquad \text{and}\qquad V_i=0. 
\end{equation}
Thus, at the linearized level, we confirm that the gravitational contribution consists only of the two tensor modes, which are given by: 
\begin{equation}
    \mathcal{L}_{EH}=-\frac{M_{pl}^2}{8}\partial_{\mu}h_{ij}^T\partial^{\mu}h_{ij}^T. 
\end{equation}

Let us now look at the vector part. Similarly to the metric perturbations, we will separate the temporal and spatial components of the vector field, and further decompose the spatial part according to the following: 
\begin{equation}\label{vecdec}
    A_i=A_i^T+\chi_{,i},\qquad A_{i,i}^T=0. 
\end{equation}
Then, the Lagrangian density corresponding to the vector field becomes: 
\begin{equation}
      \begin{split}
          \mathcal{L}_{P}=\frac{1}{2}&\left[A_0\left(-\Delta+m^2\right)A_0+2A_0\Delta\dot{\chi}-\left(\dot{\chi}\Delta\dot{\chi}-m^2\chi\Delta\chi\right)\right.\\ &\left.+\left(\dot{A_i^T}\dot{A_i^T}-A_{i,j}^TA_{i,j}^T-m^2A_i^TA_i^T\right)\right]
      \end{split}
\end{equation}
We can see that the temporal component is not-propagating, and satisfies the constraint: 
\begin{equation}
     (-\Delta+m^2)A^0=-\Delta\dot{\chi},
\end{equation}
Substituting its solution 
\begin{equation}
    A_0=\frac{-\Delta}{-\Delta+m^2}\dot{\chi}.
\end{equation}
back to the action, we find: 
\begin{equation}\label{eq::Paction3}
         \mathcal{L}_{P}=-\frac{1}{2}\left[A_i^T(-\Box+m^2)A_i^T+\chi(-\Box+m^2)\frac{m^2(-\Delta)}{-\Delta+m^2}\chi\right].
\end{equation}
Thus, we have found two transverse vector modes, $A_i^T$, and a longitudinal mode. 

At the linearized level, the \textit{dof} of the de Broglie-Proca theory are decoupled. However, once one considers higher-order contributions, that arise due to the coupling of the Proca theory to gravity, as well as the non-minimal coupling, the modes can influence each other's behavior. In this work, our aim is to investigate these corrections, and ask -- \textit{At which scale can the perturbation theory break down?} 

The main tool that will help us to answer this question is the minimal amplitude of quantum fluctuations -- equal time, two-point colleration function, which is the direct consequence of Heisenberg's uncertainty principle. For a normalized quantum field, the typical amplitude of the quantum fluctuations for scales $k$ corresponding to length scales $L\sim\frac{1}{k}$ is of the order of $\left(\frac{k^3}{\omega_k}\right)^{\frac{1}{2}}$ \cite{Mukhanov:2007zz}. 

By canonically normalizing the tensor modes according to: 
\begin{equation}
    h_{(n)ij}^T=\frac{M_{pl}}{2}h_{ij}^T,
\end{equation}
we thus find that quantum fluctuations of the original modes behave as: 
\begin{equation}\label{qflucttens}
    \delta h^T_{L}\sim\frac{1}{M_{pl}L}. 
\end{equation}

Similarly, the longitudinal modes need to be canonically normalized as well, according to: 
\begin{equation}
    \chi_n=\sqrt{\frac{-\Delta m^2}{-\Delta+m^2}}\chi,
\end{equation}
while the transverse modes are already in the canonical form. 
For scales $k^2\gg m^2$, we thus find that the minimal amplitude of quantum fluctuations for these modes is respectively given by: 
\begin{equation}\label{qfluctvec}
    \delta \chi_{L}\sim\frac{1}{mL}\qquad \text{and}\qquad \delta A_{L}^T\sim\frac{1}{L}. 
\end{equation}

In contrast to the transverse modes, we can notice that the quantum fluctuations of the longitudinal ones are singular in mass. In the presence of the self-interactions of the Proca field, it is well known that these introduce a breakdown in the perturbation theory \cite{Hell:2021wzm}. Let us now investigate if the same will hold true in the presence of the non-minimal coupling. 

\section{  {\textcolor{Black}{\Large \textbf{\textsc{Investigating the Strong Coupling}}}}}

In the previous section, we have seen that each type of mode behaves differently. In contrast to the transverse modes, which are only dependent on the length scale, the amplitude of the quantum fluctuations for the tensor modes is inversely proportional to the Planck mass. Most importantly, the amplitude of quantum fluctuations for the longitudinal modes is singular in the mass of the vector field. 

The main focus of this section will be to investigate the massless limit of the Proca theory, with non-minimal coupling to the Ricci tensor and Ricci scalar given in (\ref{PdeBroglieAction}). Due to their amplitude of quantum fluctuations, we might expect that the longitudinal modes might give rise to the perturbative series that is singular in mass. This is the case in the presence of self-interactions such as $\left(A_{\mu}A^{\mu}\right)^2$, where the longitudinal modes consequently enter a strong coupling regime \cite{Hell:2021wzm} once the non-linear terms become of the same order as the linear ones (see also \cite{Hell:2021oea} for the non-Abelian generalization). 

To clearly see the structure of each interaction, we will consider the Ricci scalar and the Ricci tensor type couplings separately, without any loss of generality. To find the behavior of the leading order corrections, we will follow the procedure introduced \cite{Chamseddine:2018gqh}, going beyond the linear analysis and estimating the corrections to each of the modes based on the minimal amplitude of the quantum fluctuations. 

By expanding the metric perturbations around the flat background as in the previous chapter, we find the following Lagrangian density: 

\begin{equation}
\mathcal{L}=\mathcal{L}_{EH}+\mathcal{L}_P+\mathcal{L}_{\alpha int}+\mathcal{L}_{\beta int},
\end{equation}
where 
\begin{equation*}
    \begin{split}
        \mathcal{L}_{EH}=&
        \frac{M_{pl}^2}{4}\left(2h^{\mu\nu}_{,\mu}h_{\alpha\nu}^{,\alpha}-h_{\mu\nu,\alpha}h^{\mu\nu,\alpha}-2h^{\mu\nu}_{,\mu}h_{,\nu}+h_{,\mu}h^{,\mu}\right)
        \\
        &+\frac{M_{pl}^2}{2}\left(-\frac{1}{4}h^{\alpha\beta}h_{\mu\nu}h^{\mu\nu}_{,\alpha\beta}-\frac{1}{8}h_{\mu\nu}h^{\mu\nu}h^{\alpha\beta}_{,\alpha\beta}-\frac{1}{4}h^{\alpha\beta}h_{\beta\nu}h_{\alpha,\delta}^{\nu,\delta}-h^{\alpha\beta}h_{\alpha,\mu}^{\mu}h^{\nu}_{\beta,\nu}\right.\\&\left. 
        +\frac{1}{2}h^{\alpha\beta}h_{\beta\nu,\mu}h^{\mu,\nu}_{\alpha}-h^{\alpha\beta}_{,\mu}h_{\alpha}^{\mu}h^{\nu}_{\beta,\nu}+\frac{1}{8}hh_{\mu\nu}h^{\mu\nu,\gamma}_{,\gamma}+\frac{3}{16}h_{\mu\nu}h^{\mu\nu}h_{,\gamma}^{,\gamma}+h^{\alpha\beta}h_{\alpha,\mu}^{\mu}h_{,\beta}\right.\\&\left.+\frac{1}{2}hh^{\mu\nu}_{,\mu}h^{\alpha}_{\nu,\alpha}-\frac{1}{4}hh_{\alpha\nu,\mu}h^{\alpha\mu,\nu}+\frac{1}{4}hh^{\alpha\beta}h_{,\alpha\beta}-\frac{1}{16}h^2h^{,\mu}_{,\mu}\right),
        \\ \mathcal{L}_P=&-\frac{1}{4}F_{\mu\nu}F^{\mu\nu}-\frac{m^2}{2}A_{\mu}A^{\mu}\\
        &+\frac{1}{2}h^{\mu\alpha}F_{\mu\nu}F_{\alpha}^{\;\nu}+\frac{m^2}{2}h^{\mu\nu}A_{\mu}A_{\nu}+\frac{1}{2}h\left(-\frac{1}{4}F_{\mu\nu}F^{\mu\nu}-\frac{m^2}{2}A_{\mu}A^{\mu}\right),
        \\
        \mathcal{L}_{\alpha int}=&-\frac{\alpha}{2}\left(h^{\mu\nu}_{,\mu\nu}-\Box h\right)A_{\mu}A^{\mu},\\
         \mathcal{L}_{\beta int}=&-\frac{\beta}{4}A_{\mu}A_{\nu}\left(2h^{\nu\alpha,\mu}_{,\alpha}-h^{,\mu\nu}-\Box h^{\mu\nu}\right)
    \end{split}
\end{equation*}

In the following analysis, we will decompose the metric and the vector field according to the spatial rotations given in (\ref{decomposition}) and \ref{vecdec}) respectively, and apply the conformal gauge $E=B=0$, together with $F_i=0$. 

In order to gain insights into the structure of the interactions, we will study each non-minimal coupling case separately. 

\subsection{ $\beta$ \textsc{ coupling case} }

As a first step, let us set $\alpha=0$, and consider only the coupling of the vector fields with the Ricci tensor. In this case, the most important terms in the Lagrangian density are given by: 

\begin{equation}\label{Ldecompolddecomp}
\mathcal{L}=\mathcal{L}_{EH}+\mathcal{L}_P+\mathcal{L}_{\beta int},
\end{equation}
where 
\begin{equation*}\begin{split}\mathcal{L}_{EH}\sim&M_{pl}^2\left(2\phi\Delta\psi-3\dot{\psi}\dot{\psi}-\psi\Delta\psi\right)-\frac{M_{pl}^2}{4}
S_i\Delta S_i-\frac{M_{pl}^2}{8}\partial_{\mu}h_{ij}^T\partial^{\mu}h_{ij}^T\\&
-\frac{M_{pl}^2}{4}\left(\frac{1}{2}h_{ij}^Th_{ik}^Th_{jk,\mu}^{T,\mu}+\frac{1}{2}h_{ij}^Th_{kl}^Th_{kl,ij}^T-h_{ij}^Th_{il,k}^Th_{jk,l}^T\right),
\end{split}
\end{equation*}
\begin{equation*}
    \begin{split}
\mathcal{L}_{P}=&\frac{1}{2}\left[A_0\left(-\Delta+m^2\right)A_0+2A_0\Delta\dot{\chi}-\left(\dot{\chi}\Delta\dot{\chi}-m^2\chi\Delta\chi\right)\right.\\ &\left.+\left(\dot{A_i^T}\dot{A_i^T}-A_{i,j}^TA_{i,j}^T-m^2A_i^TA_i^T\right)\right]\\&+\psi\left[\left(A_{j,i}^T-A_{i,j}^T\right)\left(A_{j,i}^T-A_{i,j}^T\right)-\left(\dot A_i-A_{0,i}\right)\left(\dot A_i-A_{0,i}\right)+m^2A_iA_i\right]\\
          &+\frac{1}{2}(3\psi-\phi)\left[\dot A_i\dot A_i-A_{j,i}A_{j,i}+A_{j,i}A_{i,j}+A_{0,i}A_{0,i}-2A_{0,i}\dot A_i+m^2A_0^2-m^2A_iA_i\right]\\
          &+\phi\left[\left(\dot A_i-A_{0,i}\right)\left(\dot A_i-A_{0,i}\right)+m^2A_0^2\right]-S_i\left[\left(\dot A_j-A_{0,j}\right)\left(A^T_{j,i}-A^T_{i,j}\right)+m^2A_0A_i\right]\\
          &+\frac{1}{2}h_{ij}^T\left[\left(A_{k,i}^T-A_{i,k}^T\right)\left(A_{k,j}^T-A_{j,k}^T\right)-\left(A_{0,i}-\dot A_i\right)\left(A_{0,j}-\dot A_j\right)+m^2A_iA_j\right]\\\\
          \mathcal{L}_{\beta int}=&\frac{\beta}{2}\left[A_0^2\left(\Delta\phi+3\Ddot{\psi}\right)-A_0A_i\left(\Delta S_i+4\dot{\psi}_{,i}\right)\right]\\
        &+\frac{\beta}{4}A_iA_j\left(2\psi_{,ij}+2\Box\psi \delta_{ij}-2\phi_{,ij}+\dot{S}_{j,i}+\dot{S}_{i,j}+\Box h_{ij}^T\right)
     \end{split}
\end{equation*}
Here, we have kept $A_i$ in some parts in order to keep expressions simpler.

One can notice that we have neglected the cubic terms of $\psi$ ad $S_i$, as well as their combinations that involve $\phi$ in $\mathcal{L}_{EH}$. These fields would yield at least quartic terms that mix with the tensor modes. The reason for this lies in the constraints. By varying with respect to $\phi$, one finds that $\psi$ satisfies to the leading order:  
\begin{equation}
    \begin{split}
        \psi=&\frac{\beta}{4M_{pl}^2}\left[\frac{\partial_i\partial_j}{\Delta}\left(A_iA_j\right)-A_0A_0\right]-\frac{1}{2M_{pl}^2}\frac{1}{\Delta}\left[\left(\dot A_i-A_{0,i}\right)\left(\dot A_i-A_{0,i}\right)+m^2A_0^2\right]\\
        &+\frac{1}{4M_{pl}^2}\frac{1}{\Delta}\left[\dot A_i\dot A_i-A_{j,i}A_{j,i}+A_{j,i}A_{i,j}+A_{0,i}A_{0,i}-2A_{0,i}\dot A_i+m^2A_0^2-m^2A_iA_i\right]
    \end{split}
\end{equation}
The vector mode $S_i$ is constrained as well. The leading order solution for its constraint is given by: 
\begin{equation}
    \begin{split}
        S_k=&\frac{\beta}{M_{pl}^2}P_{ki}^T\left[-A_0A_i+\frac{1}{2}\frac{1}{\Delta}\partial_j\partial_0\left(A_jA_j\right)\right]-\frac{2}{M_{pl}^2}P_{ki}^T\left[\left(\dot A_j-A_{0,j}\right)\left(A^T_{j,i}-A^T_{i,j}\right)-m^2A_0A_i\right],
    \end{split}
\end{equation}
where 
\begin{equation}
    P_{ij}^T=\delta_{ij}-\frac{\partial_i\partial_j}{\Delta}
\end{equation}
is the transverse projector. 

By substituting these expressions back to (\ref{Ldecompolddecomp}), we find up to cubic order: 
\begin{equation}\label{Ldecompolddecomp2}
    \mathcal{L}=\mathcal{L}_{EH}+\mathcal{L}_P+\mathcal{L}_{\beta int},
\end{equation}
where 
\begin{equation*}
    \begin{split}
\mathcal{L}_{EH}=&-\frac{M_{pl}^2}{8}\partial_{\mu}h_{ij}^T\partial^{\mu}h_{ij}^T
-\frac{M_{pl}^2}{4}\left(\frac{1}{2}h_{ij}^Th_{ik}^Th_{jk,\mu}^{T,\mu}+\frac{1}{2}h_{ij}^Th_{kl}^Th_{kl,ij}^T-h_{ij}^Th_{il,k}^Th_{jk,l}^T\right),\\\\
\mathcal{L}_{P}=&\frac{1}{2}\left[A_0\left(-\Delta+m^2\right)A_0+2A_0\Delta\dot{\chi}-\left(\dot{\chi}\Delta\dot{\chi}-m^2\chi\Delta\chi\right)\right.\\ &\left.+\left(\dot{A_i^T}\dot{A_i^T}-A_{i,j}^TA_{i,j}^T-m^2A_i^TA_i^T\right)\right]\\&+\frac{1}{2}h_{ij}^T\left[\left(A_{k,i}^T-A_{i,k}^T\right)\left(A_{k,j}^T-A_{j,k}^T\right)-\left(A_{0,i}-\dot A_i\right)\left(A_{0,j}-\dot A_j\right)+m^2A_iA_j\right]\\\\
          \mathcal{L}_{\beta int}=&\frac{\beta}{4}A_iA_j\Box h_{ij}^T
     \end{split}
\end{equation*}
We can notice that similarly to the free case, the temporal component of the vector field is not propagating. By varying the action with respect to it, we find: 
\begin{equation}
    (-\Delta+m^2)A_0=-\Delta\dot\chi-h_{ij}^T\left(A_{0,ij}-\dot A_{j,i}\right)
\end{equation}
By solving it perturbatively, and substituting the solution back to the action, we find: 
\begin{equation}
   \begin{split}
        \mathcal{L}\sim&-\frac{M_{pl}^2}{8}\partial_{\mu}h_{ij}^T\partial^{\mu}h^{T}_{ij}-\frac{1}{2}A_i^T(-\Box+m^2)A_i^T-\frac{1}{2}\chi(-\Box+m^2)\frac{m^2(-\Delta)}{-\Delta+m^2}\chi\\
        &+\frac{\beta}{4}\left(A_i^T+\chi_{,i}\right)\left(A_j^T+\chi_{,j}\right)\Box h_{ij}^T+\frac{1}{2}h_{ij}^T\left[\left(A_{k,j}^T-A_{j,k}^T\right)\left(A_{k,i}^T-A_{i,k}^T\right)-\dot A_i^T\dot A_j^T\right]\\
        &\frac{m^2}{2}h_{ij}^T\left(\chi_{,i}\chi_{,j}+2\chi_{,i}A_j^T\right),
   \end{split}
\end{equation}
where have kept the most relevant terms.

Let us now study the equations of motion of the propagating modes. The most important contributions to the equation of motion of the longitudinal modes are given by: 
\begin{equation}\label{longold}
    (-\Box+m^2)m^2\chi\sim-\frac{\beta}{2}\partial_i\left[\chi_{,j}\Box h_{ij}^T\right]-m^2\partial_i\left(h_{ij}^T\chi_{,j}\right)
\end{equation}
For the transverse modes, we find: 
\begin{equation}
    \begin{split}
        (-\Box+m^2)A_l^T=P_{li}^T&\left[\frac{\beta}{2}\left(A_j^T+\chi_{,j}\right)\Box h_{ij}^T+\dot h_{ij}^T\dot A_j^T\right.\\&\left.-h_{ij}^T\left(A_{i,jk}^T-A_{j,ik}^T\right)+\partial_k\left(h_{ij}^T\left(A_{k,j}^T-A_{j,k}^T\right)\right)\right]
    \end{split}
\end{equation}
Finally, for the transverse modes, the leading order contributions are given by: 
\begin{equation}\label{tenold}
    \Box h_{ij}^T\sim -\frac{\beta}{M_{pl}^2}P_{ijkl}^T\Box\left(\chi_{,k}\chi_{,l}\right),
\end{equation} 
where 
\begin{equation}
   \begin{split}
        P_{mnij}^T&=\frac{1}{2}\left(\delta_{im}\delta_{jn}+\delta_{jm}\delta_{in}\right)-\frac{1}{2}\delta_{ij}\left(\delta_{mn}-\frac{\partial_k\partial_l}{\Delta}\delta_{km}\delta_{ln}\right)\\
        &+\frac{1}{\Delta}\left[\frac{1}{2}\delta_{mn}\partial_i\partial_j+\frac{1}{2}\delta_{mk}\delta_{ln}\frac{\partial_i\partial_j\partial_k\partial_l}{\Delta}-\delta_{im}\delta_{ln}\partial_l\partial_j-\delta_{jm}\delta_{ln}\partial_l\partial_i\right]
   \end{split}
\end{equation}
is the transverse-traceless projector.
Let us evaluate the above equations perturbatively:
\begin{equation}
    \chi=\chi^{(0)}+\chi^{(1)}+...\qquad A_i^T=A_i^{T(0)}+A_i^{T(1)}+...\qquad h_{ij}^T=h_{ij}^{T(0)}+h_{ij}^{T(1)}+...,
\end{equation}
with 
\begin{equation}
    (-\Box+m^2)\chi^{(0)}=0\qquad  (-\Box+m^2)A_i^{T(0)}=0\qquad  \Box h_{ij}^{T(0)}=0. 
\end{equation}
In this case, the first-order corrections for the longitudinal modes are given by:
\begin{equation}
     (-\Box+m^2)m^2\chi^{(1)}=-m^2\partial_i\left(h_{ij}^{T(0)}\chi_{,j}^{(0)}\right)
\end{equation}
By taking into account (\ref{qfluctvec}) and (\ref{qflucttens}), for scales $k^2\sim\frac{1}{L^2}\gg m^2$, we can evaluate them as 
\begin{equation}
    \chi^{(1)}\sim\frac{1}{m M_{pl}L^2}. 
\end{equation}
They become of the same order as the linear terms $\chi^{(0)}$ at the Planck scale: 
\begin{equation}
    L_{\chi}\sim\frac{1}{M_{pl}}. 
\end{equation}
Similarly, the first-order corrections to the transverse modes can be evaluated as: 
\begin{equation}
    A_i^{T(1)}\sim \frac{1}{M_{pl}L^2},
\end{equation}
and become of the same order as the linear term at the same scale as the longitudinal mode.

In contrast to the transverse and the longitudinal modes, we can notice that the first-order corrections, which can be estimated as: 
\begin{equation}
    h_{ij}^{(1)}\sim\frac{\beta}{M^2_{pl}m^2L^4},
\end{equation}
are singular in mass. Once they become of the same order as the linear term $h_{ij}^{T(0)}\sim\frac{1}{M_{pl}L}$, the tensor modes enter a strong coupling regime.
This corresponds to the following length scale:
\begin{equation}
    L_{\beta str}\sim \left(\frac{\beta}{M_{pl}m^2}\right)^{1/3},
\end{equation} 
appearing sooner than the Planck length is reached.  

As the perturbation theory is no longer applicable at scales corresponding to the previous strong coupling scale, the first term on the left-hand side of the equation for the longitudinal modes (\ref{longold}) is no longer vanishing. Once the scale $L_{\beta str}$ is reached, their first-order corrections should be evaluated rather as: 
\begin{equation}
    \chi^{(1)}\sim \frac{\beta}{m^3M_{pl}L^4}.
\end{equation}
Due to them, the longitudinal mode becomes strongly coupled as well, as soon as one reaches the scale  $L_{\beta str}$. Moreover, by studying the quartic order contributions, one can find that terms of the order $\mathcal{O}\left(\frac{\beta^2}{M_{pl}^2L^6}\chi^4\right)$ appear in the Lagrangian density, which is of the same order as the cubic contributions when evaluated at the strong coupling scale. 

When studying Proca theory, with interactions that are perturbatively singular in the limit when the mass goes to zero, it is natural to expect that the longitudinal mode becomes strongly coupled, as this is the source of the singular behavior, and is absent in the massless theory. Yet, in this subsection, we find that due to the coupling with the Ricci tensor, not only the longitudinal mode becomes strongly coupled,  but the tensor modes as well, once the scale $L_{\beta str}$ is reached. However, this is an unphysical result. It would imply that no matter how small the mass of the photon is, gravitational waves would be strongly coupled if non-minimal coupling with the Ricci tensor is taken into account.

Let us now study the coupling with the Ricci scalar, and evaluate if the same effect takes place. 

\subsection{ $\alpha$ \textsc{ coupling case} }
In the previous section we have seen that due to the $\beta$ interaction, both tensor and the longitudinal modes become strongly coupled. Let us now study the coupling that involves only the Ricci scalar. We can notice that at cubic order, only the gravitational potentials are included in the interaction: 
\begin{equation}
    \mathcal{L}_{\alpha int}= -\alpha\left(\Delta\phi+3\Ddot{\psi}-2\Delta\psi\right)A_{\mu}A^{\mu}
\end{equation}
Therefore, in contrast to the previous case, we can safely avoid the danger of tensor modes becoming strongly coupled in this order. Thus, the only question to be answered is at which scale will the longitudinal modes enter a strong coupling regime. For simplicity, let us set the tensor and transverse vector modes to zero. Then, the most dominant terms in the Lagrangian are given by:
\begin{equation}
    \mathcal{L}=\mathcal{L}_{EH}+\mathcal{L}_P+\mathcal{L}_{\alpha int},
\end{equation}
where 
\begin{equation}
    \begin{split}
        \mathcal{L}_{EH}=&M_{pl}^2\left(2\phi\Delta\psi-3\dot{\psi}\dot{\psi}-\psi\Delta\psi\right)\\
        \mathcal{L}_{P}=&\frac{1}{2}\left[A_0\left(-\Delta+m^2\right)A_0+2A_0\Delta\dot{\chi}-\left(\dot{\chi}\Delta\dot{\chi}-m^2\chi\Delta\chi\right)\right]\\
         \mathcal{L}_{\alpha int}=& -\alpha\left(\Delta\phi+3\Ddot{\psi}-2\Delta\psi\right)\left(\chi_{,i}\chi_{,i}-A_0A_0\right)
    \end{split}
\end{equation}
One should note that we have neglected also higher orders of the gravitational potentials as they would generate terms that are in higher order than quartic ones. Moreover, we have neglected the interactions that arise from the $\mathcal{L}_P$, as these will be multiplied by the mass term after resolving the constraints, and thus be subdominant. 

Let us now study the equations of motion. By varying with respect to $\phi$, we find:
\begin{equation}
    2M_{pl}^2\Delta\psi=\alpha\Delta\left(\chi_{,i}\chi_{,i}-A_0A_0\right),
\end{equation}
which can be solved as: 
\begin{equation}\label{psialpha}
    \psi=\frac{\alpha}{2M_{pl^2}}\left(\chi_{,i}\chi_{,i}-A_0A_0\right).
\end{equation}
By varying with respect to $\psi$, we further find: 
\begin{equation}
    2M_{pl}\Delta\phi+6M_{pl}^2\Ddot{\psi}-2M_{pl}^2\Delta\psi-3\alpha\partial_0^2\left(\chi_{,i}\chi_{,i}-A_0A_0\right)+2\alpha\Delta\left(\chi_{,i}\chi_{,i}-A_0A_0\right)=0
\end{equation}
By substituting (\ref{psialpha}) into the above equation, we find:
\begin{equation}\label{psiisphialpha}
    \phi=-\psi
\end{equation}
within our approximation. Let us now consider the $A_0$ constraint. By varying the action with respect to this component, we find: 
\begin{equation}
    (-\Delta+m^2)A_0-6\alpha\Box\psi=-\Delta\dot\chi,
\end{equation}
where we have used (\ref{psiisphialpha}). In the limit $k^2\gg m^2$, the solution of the above constraint is given by: 
\begin{equation}
    A_0\sim\frac{-\Delta}{-\Delta+m^2}\dot\chi+\frac{3\alpha^2}{M_{pl}^2}\frac{1}{-\Delta+m^2}\left[\dot\chi\Box(\chi_{,i}\chi_{,i}-\dot{\chi}\dot \chi)\right]
\end{equation}
Finally, the longitudinal mode satisfies the following equation of motion: 
\begin{equation}
    \Delta\Ddot{\chi}+m^2\Delta\chi-\Delta\dot A_0-6\alpha\partial_i\left(\chi_{,i}\Box\psi\right)=0,
\end{equation}
where we have used (\ref{psiisphialpha}) again. Substituting in addition the solution for the constraint of the temporal component, this equation can be written on scales $k^2\gg m^2$ as: 
\begin{equation}
    -m^2(-\Box+m^2)\chi\sim\frac{3\alpha^2}{M_{pl}^2}\partial_{\mu}\left[\chi^{,\mu}\Box(\chi_{,\alpha}\chi^{,\alpha})\right]. 
\end{equation}
The term on the right-hand side is not vanishing even within the perturbation theory. By taking into account the minimal amplitude of quantum fluctuations for the longitudinal mode, we can notice that it is also singular in mass. It will become of the same order as the linear term $\chi^{(0)}$ at the strong coupling scale:
 \begin{equation}
     L_{\alpha}\sim\left(\frac{\alpha}{M_{pl}m^2}\right)^{1/3}
 \end{equation}
We can notice that this scale arrives in the same manner as the strong coupling scale of the $\beta$ case, with only the coupling interchanged. However, while for the $\beta$ case we found that both tensor and longitudinal modes become strongly coupled, here only the longitudinal modes enter the strong coupling regime. This is also to be expected -- the longitudinal modes are absent in the massless theory, while the tensor and transverse modes remain there. Moreover, due to the scaling of the remaining modes -- the vector and tensor ones -- we can be sure that this will be the most dominant term. 

\section{  {\textcolor{Black}{\Large \textbf{\textsc{How to cure the tensor modes? }}}}}

In the previous sections, we have studied the $\alpha$ and $\beta$ models, which involved the coupling of the vector field with the Ricci scalar and the Ricci tensor respectively. By writing the theory only in terms of the propagating modes, and by studying their corresponding equations of motion, we have shown that within the perturbation theory, both types of coupling are seemingly singular in mass, and consequently introduce strong coupling scales. In the case of the Ricci scalar, we find that only the longitudinal mode becomes strongly coupled, due to the quartic self-interaction that appears after one resolves all of the constraints. In contrast, the coupling of the vector field with the Ricci tensor indicates that both the longitudinal mode and the tensor mode become strongly coupled, leading us to a physical ambiguity. 

On the one hand, one might simply discard the couplings with the Ricci tensor and focus only on the coupling to the Ricci scalar as a viable coupling. In this work, we will, however, try to resolve this puzzle in another way. 

Interestingly, a similar problem appears in the case of massive Yang-Mills theory, with mass added by hand \cite{Hell:2021oea}. In flat space-time, this theory describes the massive SU(2) vector field $A_{\mu}$ and is given by the action: 
\begin{equation}\label{eq::action}
    S=\int d^4x \left[-\frac{1}{2}\text{Tr}(F_{\mu\nu}F^{\mu\nu})+m^2\text{Tr}(A_{\mu}A^{\mu})\right],
\end{equation}
where 
\begin{equation}
      F_{\mu\nu}=D_{\mu}A_{\nu}-D_{\nu}A_{\mu},\qquad \text{and}\qquad D_{\mu}=\partial_{\mu}+igA_{\mu}
\end{equation}
are the field strength tensor and the covariant derivative respectively, while  $g$ is the coupling constant. Similarly to the Proca theory, the temporal component of the vector field is not propagating. If one finds its constraint, solves it, and substitutes it back to the action, one will find the action only in terms of the transverse and the longitudinal modes. Similarly to (\label{vecdec}), we can decompose the spatial part of the vector field as: 
\begin{equation}\label{mYMdec}
    A_i^a=A_i^{Ta}+\chi_{,i},\qquad\text{where}\qquad A_{i,i}^{Ta}=.
\end{equation}
Here, we have expressed the non-Abelian vector field in terms of the generators of the SU(2) group 
\begin{equation}
    A_{\mu}=A_{\mu}^aT^a, \qquad a=1,2,3,    
\end{equation}
where $T^a=\frac{\sigma^a}{2}$ are the Pauli matrices. However, in the case of the decomposition (\ref{mYMdec}), one will find that the transverse modes enter the strong coupling regime in addition to the strong coupling of the longitudinal modes, in a way that is similar to the tensor modes in the $\beta$ coupling case:
\begin{equation}\label{eq::Translineom}
    (\Box+m^2)A_i^{Ta}\sim g\varepsilon^{abc}P_{ij}^T\left[\frac{1}{2}\Box\left(\chi^b\chi^c_{,j}\right)\right],\qquad  P_{ij}^T=\delta_{ij}-\frac{\partial_i\partial_j}{\Delta}
\end{equation}
The right-hand side of the above equation, however, suggests that the introduction of the coupling in the form of such a term could remove the strong coupling of the transverse modes. Notably, by introducing the coupling in the following way, 
\begin{equation}\label{newcouplingmYM}
    \begin{split}
        A_i^a=A_i^{Ta}+\chi_{,i}^a-g\varepsilon^{abc}(\frac{1}{2}\chi_{,i}^b\chi^c),
    \end{split}
\end{equation}
one not only removes the above strong coupling of the transverse modes but also shifts the strong coupling scale. In fact, by studying the new theory with the coupling given by (\ref{newcouplingmYM}), one will still find unphysical results for the transverse modes, now at a higher scale, and should repeat the procedure by introducing an infinite number of couplings, defined in the following way: 
\begin{equation}\label{eq::NonlindecmYM}
    A_i=\zeta A_i^T\zeta^{\dagger}+\frac{i}{g}\zeta_{,i}\zeta^{\dagger}\qquad\text{with}\qquad   A_{i,i}^{T}=, 
\end{equation}
where $\zeta$ is the unitary matrix given by
\begin{equation}
    \zeta=e^{-ig\chi}.
\end{equation}
In this case, only the longitudinal modes will become strongly coupled, with the strong coupling scale that matches the scale at which the unitarity is violated. 

Following the case for the mYM theory, we could introduce additional coupling to the theory such that the strong coupling of the tensor mode disappears. Following the procedure found in \cite{Hell:2021oea}, we can notice that at the strong coupling scale $L_{\beta str}$, both terms are of the same order in the equation of motion for the tensor modes: 
\begin{equation}
    \Box h_{ij}^T\sim -\frac{\beta}{M_{pl}^2}P_{ijkl}^T\Box\left(\chi_{,k}\chi_{,l}\right),
\end{equation} 
The form of this equation suggests that the tensor modes should be transformed: 
\begin{equation}
    h_{ij}^T= \Bar{h}_{ij}^T-\frac{\beta}{M_{pl}^2}P_{ijkl}^T\left(\chi_{,k}\chi_{,l}\right).
\end{equation}

This form of the introduction of new couplings, known as the disformal couplings suggests that writing the metric perturbations in a manifestly covariant way as:
\begin{equation}\label{newdecomposition}
    g_{\mu\nu}=\eta_{\mu\nu}+h_{\mu\nu}-\frac{\beta}{M_{pl}^2}A_{\mu}A_{\nu},
\end{equation}
could remove the problematic coupling between the tensor and longitudinal modes. 
Notably, this form could be also deducted directly from the action. For this, one should note that the Ricci tensor vanishes for the flat background, thus having the dependence only on the perturbations. Therefore, one can trace the appearance of the tensor modes among them, and then by looking at the form of the coupling infer the resulting disformal transformation.

In the remaining part of this section, we will study the disformal theory, and investigate if the disformal coupling removes the strong coupling of the tensor modes. In addition, we will have to make sure if the coupling previously introduced is the only necessary coupling, or similarly to the mYM case one has to introduce multiple ones. Finally, we will verify if the longitudinal modes nevertheless enter the strong coupling regime, and if the resulting strong coupling scale differs from the one that we have initially found. To answer these questions, let us study the longitudinal modes at the 
quartic order.

With (\ref{newdecomposition}), we find the following Lagrangian density: 
\begin{equation}
\mathcal{L}=\mathcal{L}_{h}+\mathcal{L}_{A}+\mathcal{L}_{Ah}
\end{equation}
\begin{equation*}
    \begin{split}
        \mathcal{L}_{h}=&
        \frac{M_{pl}^2}{4}\left(2h^{\mu\nu}_{,\mu}h_{\alpha\nu}^{,\alpha}-h_{\mu\nu,\alpha}h^{\mu\nu,\alpha}-2h^{\mu\nu}_{,\mu}h_{,\nu}+h_{,\mu}h^{,\mu}\right)
        \\
        &+\frac{M_{pl}^2}{2}\left(-\frac{1}{4}h^{\alpha\beta}h_{\mu\nu}h^{\mu\nu}_{,\alpha\beta}-\frac{1}{8}h_{\mu\nu}h^{\mu\nu}h^{\alpha\beta}_{,\alpha\beta}-\frac{1}{4}h^{\alpha\beta}h_{\beta\nu}h_{\alpha,\delta}^{\nu,\delta}-h^{\alpha\beta}h_{\alpha,\mu}^{\mu}h^{\nu}_{\beta,\nu}\right.\\&\left. 
        +\frac{1}{2}h^{\alpha\beta}h_{\beta\nu,\mu}h^{\mu,\nu}_{\alpha}-h^{\alpha\beta}_{,\mu}h_{\alpha}^{\mu}h^{\nu}_{\beta,\nu}+\frac{1}{8}hh_{\mu\nu}h^{\mu\nu,\gamma}_{,\gamma}+\frac{3}{16}h_{\mu\nu}h^{\mu\nu}h_{,\gamma}^{,\gamma}+h^{\alpha\beta}h_{\alpha,\mu}^{\mu}h_{,\beta}\right.\\&\left.+\frac{1}{2}hh^{\mu\nu}_{,\mu}h^{\alpha}_{\nu,\alpha}-\frac{1}{4}hh_{\alpha\nu,\mu}h^{\alpha\mu,\nu}+\frac{1}{4}hh^{\alpha\beta}h_{,\alpha\beta}-\frac{1}{16}h^2h^{,\mu}_{,\mu}\right),
    \end{split}
\end{equation*}
\begin{equation*}
    \begin{split}
        \mathcal{L}_{A}=&-\frac{1}{4}F_{\mu\nu}F^{\mu\nu}-\frac{m^2}{2}A_{\mu}A^{\mu}\\
        &+\frac{\beta^2}{8M_{pl}^2}\left[\partial_{\mu}\left(A_{\alpha}A_{\beta}\right)\partial^{\mu}\left(A^{\alpha}A^{\beta}\right)+\partial_{\mu}\left(A_{\alpha}A^{\alpha}\right)\partial^{\mu}\left(A_{\beta}A^{\beta}\right)-2\partial_{\mu}\left(A^{\mu}A^{\nu}\right)\partial_{\alpha}\left(A^{\alpha}A_{\nu}\right)\right]
    \end{split}
\end{equation*}
and 
\begin{equation*}
   \begin{split}
       \mathcal{L}_{hA}=&-\frac{\beta}{4}A_{\gamma}A^{\gamma}\left(h^{\mu\nu}_{,\mu\nu}-\Box h\right)\\
       &+\frac{1}{2}h^{\mu\alpha}F_{\mu\nu}F_{\alpha}^{\;\nu}+\frac{m^2}{2}h^{\mu\nu}A_{\mu}A_{\nu}+\frac{1}{2}h\left(-\frac{1}{4}F_{\mu\nu}F^{\mu\nu}-\frac{m^2}{2}A_{\mu}A^{\mu}\right)
   \end{split}
\end{equation*}

Let us now study the above Lagrangian density in terms of components. Our goal is to find the leading order contributions for all of the modes. As we will see, the cubic coupling between $\phi$, $\psi$, and $S_i$ that appears in $\mathcal{L}_h$ will introduce quartic order terms that are only mixed between the vector modes and the tensor ones. Since we are only interested in the longitudinal modes at quartic order, we can (safely) neglect these terms. 

Moreover, the couplings between $\phi$, $\psi$, and $S_i$ and the vector modes in the last line of $\mathcal{L}_{hA}$ can be neglected as well. One can show that the longitudinal mode in these relations enters in combination with the $A_0$ component, such that once the temporal constraint is resolved, it is multiplied by the mass squared, making such terms sub-dominant. 
Therefore, the most important terms are given by: 
\begin{equation}
   \begin{split}
       \mathcal{L}\sim&M_{pl}^2\left(2\phi\Delta\psi-3\dot{\psi}\dot{\psi}-\psi\Delta\psi\right)-\frac{M_{pl}^2}{4}
S_i\Delta S_i-\frac{M_{pl}^2}{8}\partial_{\mu}h_{ij}^T\partial^{\mu}h_{ij}^T
\\&
-\frac{M_{pl}^2}{4}\left(\frac{1}{2}h_{ij}^Th_{ik}^Th_{jk,\mu}^{T,\mu}+\frac{1}{2}h_{ij}^Th_{kl}^Th_{kl,ij}^T-h_{ij}^Th_{il,k}^Th_{jk,l}^T\right)\\&
+\frac{1}{2}\left[A_0\left(-\Delta+m^2\right)A_0+2A_0\Delta\dot{\chi}-\left(\dot{\chi}\Delta\dot{\chi}-m^2\chi\Delta\chi\right)\right.\\ &\left.+\left(\dot{A_i^T}\dot{A_i^T}-A_{i,j}^TA_{i,j}^T-m^2A_i^TA_i^T\right)\right]\\&
+\frac{\beta}{4}\left(A_0^2-A_iA_i\right)\left(2\Delta\phi+6\Ddot{\psi}4-\Delta\psi\right)
+\frac{1}{2}h_{ij}^T\left[\left(A_{k,i}^T-A_{i,k}^T\right)\left(A_{k,j}^T-A_{j,k}^T\right)-\dot A^T_i\dot A^T_j\right]\\&
+\frac{\beta^2}{8M_{pl}^2}\left[\partial_{\mu}\left(A_{\alpha}A_{\beta}\right)\partial^{\mu}\left(A^{\alpha}A^{\beta}\right)+\partial_{\mu}\left(A_{\alpha}A^{\alpha}\right)\partial^{\mu}\left(A_{\beta}A^{\beta}\right)-2\partial_{\mu}\left(A^{\mu}A^{\nu}\right)\partial_{\alpha}\left(A^{\alpha}A_{\nu}\right)\right]
   \end{split}
\end{equation}

Let us now study the equations of motion. To the leading order, by varying with respect to $\phi$ and $\psi$, we find: 
\begin{equation}
    \phi\sim-\psi\sim\frac{\beta}{4M_{pl}^2}\left(A_0A_0-A_iA_i\right)
\end{equation}
and 
\begin{equation}
    S_i\sim 0. 
\end{equation}
The $A_0$ component of the vector field satisfies the following equation: 
\begin{equation}
    \begin{split}
        (-\Delta+m^2)A_0\sim&-\Delta\dot\chi-3\beta A_0\Box\psi-\frac{\beta^2}{M_{pl}^2}\left[-A_0\Delta(A_0A_0)+A_0\partial_0\partial_i(A_0\chi_{,i})\right.\\
        &\left.+\frac{1}{2}\chi_{,i}\partial_i\partial_0(A_0A_0)-\frac{1}{2}A_0\partial_0^2(\chi_{,i}\chi_{,i})+\frac{1}{2}\chi_{,i}\Delta(A_0\chi_{,i})\right.\\&\left.-\frac{1}{2}\chi_{,i}\partial_i\partial_k(A_0\chi_{,k})+\frac{1}{2}A_0\Delta(\chi_{,i}\chi_{,i})-\frac{1}{2}\chi_{,i}\partial_0\partial_j(\chi_{,i}\chi_{,j})\right]
    \end{split}
\end{equation}

By solving this equation perturbatively, and substituting it into the equation of motion for the longitudinal modes, we find that the leading corrections to the longitudinal modes satisfy: 
\begin{equation}
    \begin{split}
        m^2(-\Box+m^2)\chi^{(2)}\sim&\frac{\beta^2}{M_{pl}^2}\left[\frac{9}{2}\Delta\chi\dot\chi_{,i}\dot\chi_{,i}+11\dot\chi\dot\chi_{,i}\Delta\chi_{,i}-\dot\chi\Delta\dot\chi\Delta\chi-\dot\chi\dot\chi_{,ij}\chi_{,ij}+2\dot\chi\dot\chi\Delta^2\chi\right.\\&\left.+\dot\chi\Delta\dot\chi_{,i}\chi_{,i}+\dot\chi_{,i}\dot\chi_{,ij}\chi_{,j}-\frac{7}{4}\Delta\chi\Delta\chi\Delta\chi-\frac{7}{4}\Delta\chi\chi_{,ij}\chi_{,ij}-\frac{3}{2}\Delta\chi\Delta\chi_{,i}\chi_{,i}\right.\\&\left.
        -\frac{3}{2}\chi_{,k}\chi_{,ji}\chi_{ijk}-\chi_{,i}\Delta\chi_{,j}\chi_{,ij}-\chi_{,j}\chi_{,ik}\chi_{,jk}-\chi_{,j}\chi_{,k}\Delta\chi_{,jk}\right]
    \end{split}
\end{equation}
It can be easily shown that the first-order corrections $\chi^{(1)}$ will in this case be subleading. 
Similarly to the previous sections, we can estimate the leading order corrections to the longitudinal modes as: 
\begin{equation}
    \chi^{(2)}\sim\frac{\beta^2}{M_{pl}^2m^5L^7}
\end{equation}
These become of the same order as the linear term $\chi^{(0)}$ precisely at the strong coupling scale that we have previously found: 
\begin{equation}
    L_{\beta str}\sim\left(\frac{\beta}{M_{pl}m^2}\right)^{1/3}.
\end{equation}
At this scale, the most dominant term at the level of Lagrangian density for the longitudinal modes is of the form $\frac{\beta^2}{M_{pl}^2L^6}\chi^4$. This suggests that beyond the strongly coupled scale, the minimal amplitude of the longitudinal mode is given by: 
\begin{equation}
    \delta\chi\sim\sqrt{\frac{M_{pl}L}{\beta}}
\end{equation}
In contrast to the longitudinal modes, we can notice that the tensor and transverse modes are weakly coupled. Their leading contribution comes in the cubic order, and is of the form: 
\begin{equation}
    \begin{split}
        (-\Box+m^2)A_l^T=P_{li}^T&\left[\frac{\beta}{2}\left(A_j^T+\chi_{,j}\right)\Box h_{ij}^T+\dot h_{ij}^T\dot A_j^T\right.\\&\left.-h_{ij}^T\left(A_{i,jk}^T-A_{j,ik}^T\right)+\partial_k\left(h_{ij}^T\left(A_{k,j}^T-A_{j,k}^T\right)\right)\right]
    \end{split}
\end{equation}
for the transverse modes, and 
\begin{equation}
    M_{pl}^2\Box h_{ij}^T\sim -2(A_{k,i}^T-A_{i,k}^T)(A_{k,j}^T-A_{j,k}^T)+2\dot A_i^T\dot A_j^T+\mathcal{O}\left((h_{ij}^T)^3\frac{M_{pl}^2}{L^2}\right)
\end{equation}
for the tensor modes. 
Here, we have omitted the contribution of the longitudinal modes, which is now subdominant. By taking into account that the minimal amplitude of quantum fluctuations for the tensor modes is singular in the Planck mass, it is clear that the two types of modes will become strongly coupled only at the Planck scale, before which the perturbative approach still holds. 

Therefore, we have found that the introduction of the disformal coupling of the form (\ref{newdecomposition}) removes the strong coupling of the tensor modes. At the same time, the strong coupling of the longitudinal modes remains in the theory. Curiously, we should notice that, unlike the mYM case, the strong coupling scale does not change upon the introduction of the disformal coupling. In the following section, we will see the benefits of the introduction of the coupling (\ref{newdecomposition}) also for the cosmological background.

\section{  {\textcolor{Black}{\Large \textbf{\textsc{Resolving the Runaway Modes }}}}}
Recently, it was reported in \cite{Capanelli:2024pzd} that the non-minimally Proca theory in curved space-time has a \textit{runaway production mechanism} -- mechanism, in which at high energies, modes with negative frequency appear. Following the procedure of \cite{Capanelli:2024pzd}, let us consider again the action, with the background metric given by: 
\begin{equation}
    ds^2=a^2(\eta)\eta_{\mu\nu}dx^{\mu}dx^{\nu}. 
\end{equation}
\begin{equation}
  \begin{split}
        S_{Pnmin}=\int d^4x \sqrt{-g}&\left(-\frac{1}{4}g^{\mu\alpha}g^{\nu\beta}F_{\mu\nu}F_{\alpha\beta}-\frac{m^2}{2}g^{\mu\nu}A_{\mu}A_{\nu}\right.\\&\left.-\frac{1}{2}\alpha  Rg^{\mu\nu}A_{\mu}A_{\nu}-\frac{1}{2}\beta R^{\mu\nu}A_{\mu}A_{\nu}\right).
  \end{split}
\end{equation}
The above action can be written as: 
\begin{equation}
     S_{Pnmin}=\int d\eta d^3x\left(-\frac{1}{4}F_{\mu\nu}F^{\mu\nu}+\frac{1}{2}A_0^2M_T^2-\frac{1}{2}A_iA_iM_S^2\right),
\end{equation}
where 
\begin{equation}
    M_T^2=a^2\left[m^2+\alpha R-\beta\left(3H^2-\frac{1}{2}R\right)\right],
\end{equation}
and 
\begin{equation}
    M_S^2=a^2\left[m^2+\alpha R+\beta\left(H^2+\frac{1}{6}R\right)\right].
\end{equation}
Here, $F_{0i}=A_i'-A_{0,i}$, where the prime denotes the derivative with respect to the conformal time, and $H=\frac{a'}{a^2}$ is the Hubble parameter. 
As before, the $A_0$ component is not propagating. By varying the action with respect to it, we find the following constraint: 
\begin{equation}
    (-\Delta+M_T^2)A_0=-\Delta\chi'.
\end{equation}
By substituting its solution back to the action, and going to the momentum space for the longitudinal modes: 
\begin{equation}
    \chi(\Vec{x},\eta)=\int \frac{d^3k}{(2\pi)^{3/2}}e^{i\Vec{k}\Vec{x}}\chi_k(\eta)
\end{equation}
we find the following action for the longitudinal modes: 
\begin{equation}
   S_{\chi}=\int d\eta d^3k\frac{1}{2}\left(\frac{M_T^2k^2}{k^2+M_T^2}\chi_{-k}'\chi_k'-M_S^2k^2\chi_k\chi_{-k}\right). 
\end{equation}
By canonically normalizing the field according to: 
\begin{equation}
    \chi_{nk}=\sqrt{\frac{M_T^2k^2}{k^2+M_T^2}}\chi_k,
\end{equation}
we find, in the limit when $k^2\gg M_T^2, M_S^2$:
\begin{equation}
   S_{\chi}\sim\int d\eta d^3k\frac{1}{2}\left(\chi_{n(-k)}'\chi_{nk}'-\frac{M_S^2k^2}{M_T^2}\chi_{nk}\chi_{n(-k)}\right). 
\end{equation}
As pointed out in \cite{Capanelli:2024pzd}, if we assume that $M_T^2>0$, in order to avoid the negative kinetic term for the longitudinal modes, then, the case $M_S^2<0$ leads to the modes with negative frequency. These are called the runaway modes. In other words, with such a choice, one is able to produce modes with arbitrarily large momentum.

One should note though, that for this choice, the transverse modes have negative mass. Their Lagrangian density is given by:  
\begin{equation}
    \mathcal{L}_T=\frac{1}{2}\left(A_i^{T'}A_i^{T'}-A_{i,j}^TA_{i,j}^T-M_S^2A_i^TA_i^T\right). 
\end{equation}

We can notice that the origin of the runaway case is the $\beta$ coupling, due to which $M_T^2\neq M_S^2$. However, this is precisely the interaction between the Ricci tensor and the vector modes, that was yielding unphysical results in the flat case. Following the previous procedure which resolves the strong coupling of the tensor modes, it is thus natural to introduce the disformal coupling to the metric with the transformation that we have performed before: 
\begin{equation}
    g_{\mu\nu}=\Tilde{g}_{\mu\nu}-\frac{\beta}{M_{pl}^2}A_{\mu}A_{\nu},\qquad \text{where} \qquad \Tilde{g}_{\mu\nu}=a^2\eta_{\mu\nu}
\end{equation}
is the new metric. By performing this substitution, we find that the action (\ref{PdeBroglieAction}) becomes: 
\begin{equation}\label{disformalaction}
    \begin{split}
        S=\int d^4x\sqrt{-\Tilde{g}}&\left[\frac{M_{pl}^2}{2}\Tilde{R}-\frac{1}{4}\Tilde{g}^{\mu\alpha}\Tilde{g}^{\nu\beta}F_{\mu\nu}F_{\alpha\beta}-\frac{m^2}{2}\Tilde{g}^{\mu\nu}A_{\mu}A_{\nu}\right.\\&\left.-\frac{1}{2}(\alpha+\beta)\Tilde{R}\Tilde{g}^{\mu\nu}A_{\mu}A_{\nu}+\mathcal{O}\left(\frac{\beta^2}{M_{pl}^2}A^4\Tilde{R}\right)\right]
    \end{split}
\end{equation}
in the lowest order, while the last term represents higher-order terms that involve couplings between the vector fields with the Ricci scalar or Ricci tensor, this guarantees that now $M_S^2=M_T^2$, and thus the runaway case disappears. It should be stressed that although not explicitly analyzed, the action (\ref{disformalaction}), indicates that the theory in this disformal frame will have no strong coupling ambiguity in the tensor or runaway modes even for an arbitrary background. 

One should note that if (\ref{newdecomposition}) is taken as a field redefinition, it may happen that the instability persists in the theory, through the coupling with matter. This will certainly be interesting to check in future works. In this work, however, one should stress that (\ref{newdecomposition}) in both section 4 and this section is taken as a definition of the disformal frame, where matter should be minimally coupled to gravity. This is similar to the choice of either the Jordan or the Einstein frame in f(R) theories of gravity for the frame in which the matter is minimally coupled to gravity. In our theory, the disformal frame is the natural choice for the minimal coupling of the matter to gravity which avoids the runaway modes as well as the strong coupling of the tensor modes.

\section{  {\textcolor{Black}{\Large \textbf{\textsc{Conclusion}}}}}
In this work, we have studied Proca theory, non-minimally coupled to gravity, with two types of couplings -- one with the Ricci scalar and the Ricci tensor. The cosmological aspects of this theory have recently drawn a lot of attention, making it an intriguing possibility to underline the gravitational production of dark matter and a possible candidate for generating primordial magnetic fields and inflationary candidates.

By studying this theory in flat space-time, with metric perturbations also taken into account, we have found a somewhat surprising result when considering the massless limit. On the one hand, the interaction of the vector field with the Ricci scalar results in the strong coupling of the longitudinal modes. This is natural, as the longitudinal modes are absent in the massless theory. Thus, similarly to the Proca theory in the presence of quartic and cubic self-interactions, the longitudinal modes enter a strong coupling regime once the corresponding non-linear term becomes of the same order as the linear one. Beyond the strong coupling scale, the vector and tensor modes remain weakly coupled, up to the Planck length, due to the amplitude of quantum fluctuations for the tensor modes. 

The coupling with the Ricci tensor, on the other hand,  indicates that both tensor and longitudinal modes become strongly coupled at the same length scale. This, however, is an unphysical result -- it would imply that if one allows for such non-minimal coupling in nature, and a photon that has a small mass, the gravitational waves would quickly lose their linear propagator. Curiously, this is not the only complication that arises when the coupling with the Ricci tensor is taken into account. A recent study has found a new class of negative frequency modes in curved space-time, known as the runaway modes in the presence of non-minimal coupling \cite{Capanelli:2024pzd}.  In this work, we have shown that the origin of such modes lies in the coupling of the vector field to the Ricci tensor, which is problematic even in the flat background. One should note that the coupling with the Ricci scalar, on the other hand, does not give rise to such modes.  

Nevertheless, by carefully analyzing the theory at this strong coupling scale, we have found that the tensor modes have to be transformed. A similar scenario was found also in massive Yang-Mills theory, with mass added \textit{by hand}, where with standard decomposition, the transverse modes were becoming strongly coupled. However, by correcting their definition, it was shown that ultimately, these modes remain weakly coupled at all scales. 
In the case of non-minimally coupled Proca theory, we have found that the introduction of corresponding disformal coupling in the
theory simultaneously removes the inconsistent interaction by introducing the necessary coupling. This leads us to the self-interaction of the longitudinal modes at the quartic order, becoming a necessary step to make the theory safe. Curiously, in this disformal frame, the strong coupling scale is left unchanged, in contrast to the case of mYM theory. 

Notably, in flat spacetime, the form of the disformal coupling can also be deducted by directly studying the coupling with the Ricci tensor and the vector fields on the level of action. In this case, the Ricci tensor depends only on the perturbations, so we can trace the appearance of the interaction between the tensor modes and the longitudinal dof. In the cosmological backgrounds, this is no longer apparent, as the background value of the Ricci tensor is non-vanishing. Curiously, however, the disformal frame is defined by the same form of the disformal transformation. 
In it, the runaway issue is absent due to the disformal transformation which at the same generates further self-interactions of the vector field. It is important to stress, that in this frame the matter is minimally coupled to gravity, which guarantees that the runaway modes will not re-appear in a more complicated way. Given that, in this frame, the instability is absent (and the remnant of the coupling between the Ricci tensor remains in the higher couplings), and at the same time it resolves the strong coupling of the tensor modes in the Minkowski background, it indicates that the theory formulated in this way will be healthy also for general background.

While the disformal frame is one possible solution to the runaway and the strong coupling of tensor modes, one might regard it as fine-tuning. In a more general setting, the instability is expected to remain.  For instance, if the matter is minimally coupled to gravity in the original frame, the runaway might reappear in a non-trivial way. It would be nevertheless interesting to study this case and confirm if the instability persists, or vanishes as well.

Another interesting aspect is to study the quantized theory and to explore further aspects of the strong coupling in curved space-time for this type of theory. In particular, it would be intriguing to see if such a possibility could resolve the well-known ghost, that appears for $M_T^2<0$. 
Finally, one should note that while in Proca theory, the longitudinal modes are the source of the singularities in mass within the perturbative regime, in the Kalb-Ramond field, this role is taken by the vector modes \cite{Hell:2021wzm}. Therefore, it will be interesting to also extend this analysis to this case, and study its consequences.

\begin{center}
    \large\textsc{\textbf{Acknowledgements}}

\end{center}

I am grateful to Misao Sasaki for illuminating discussions and comments on the draft of the paper. I would also like to thank Edward "Rocky" Kolb for communicating his work on the Proca theory, and for very useful discussions. In addition, I would like to thank Elisa G. M. Ferreira, Viatcheslav Mukhanov, Shinji Mukohyama, Ryo Namba, and Ippei Obata for very useful discussions, and Evan McDonough for useful correspondence. The work of A. H. was supported by the World Premier International Research Center Initiative (WPI), MEXT, Japan.

\bibliographystyle{utphys}
\bibliography{paperbib}{}

\end{document}